\documentstyle[12pt]{article}
\newcommand{\be}{\begin{equation}}
\newcommand{\ee}{\end{equation}}
\begin{document}

\title{Ideal Gas in a Finite Container\footnote{To appear in Am. J. of Phys}}
\author{{\bf M. I.  Molina}
\vspace{1 cm}
\and
\and
Facultad de Ciencias, Departamento de F\'{\i}sica, Universidad de Chile\\
Casilla 653, Las Palmeras 3425, Santiago, Chile.}
\date{}
\maketitle
\baselineskip 24 pt
\vspace{2in}
\newpage
%%%%%%%%%%%%%%%%%%%%%%%%%%%%%%%%%%%%%%%%%%%%%%%%%%%%%%%%%%%%%%%%%%%%%
%\begin{abstract}
%\baselineskip 24 pt
%The thermodynamics of an ideal gas enclosed in a box of volume
%$a_{1} a_{2} a_{3}$ at temperature $T$ is considered. The canonical 
%partition function of the system is expressed in terms of complete elliptic
%integrals of the first kind, whose argument obeys a transcendental
%equation. For high and low temperatures we derive explicitly the
%main finite-volume corrections to the standard thermodynamic quantities.
%\end{abstract}
%%%%%%%%%%%%%%%%%%%%%%%%%%%%%%%%%%%%%%%%%%%%%%%%%%%%%%%%%%%%%%%%%%%%%
%\newpage
%%%%%%%%%%%%%%%%%% MAIN BODY %%%%%%%%%%%%%%%%%%%%%%%%%%%%%%%%%%%%%%%%%
A common practice in many statistical thermodynamics textbooks when 
calculating the partition function, is to replace the relevant 
statistical sums by integrals over a quasicontinuum 
of levels, right at the outset.$^{1}$ Since the interlevel 
separation is inversely proportional to the system's size, when doing that, 
they are really assuming an infinite system, and washing away any finite 
size effects. This, of course, simplifies the problem at hand, but at the 
same time obscures the manner in 
which size effects enter when
dealing with finite systems. In this Note, we illustrate the role of 
finite size effects by evaluating the main corrections to the canonical 
partition function of the ideal gas due to a large, yet finite, 
container volume. 

Sufficient conditions under which it is valid to replace 
sums by integrals have been examined by Stutz$^{2}$ using the
Euler summation formula. For the case of the one-particle, one-dimensional
partition function, Fox$^{3}$ rederived Stutz results by using the
Jacobi transformation. In the spirit of Fox, we will express the partition 
function of a three-dimensional ideal gas enclosed in a impenetrable box  
in terms of complete elliptic integrals of the first kind, whose argument 
obeys a transcendental equation, and will explicitly derive the main 
finite volume corrections to thermodynamical quantities. 
A different approach will be followed for the case of an impenetrable 
spherical container, where we will expand the 
density of states in powers of the energy. Surface terms appear in
the expansion coefficients, leading to finite size effects in the
thermodynamics of the system.

We will find that, for both, the box and the spherical container, the
finite volume correction terms to thermodynamical quantities, have the
{\em same} functional dependence on the ratio 
(thermal wavelength$\times$ area/volume).

\noindent{\bf Ideal gas in a box}:  Consider an ideal monoatomic 
gas of $N$ identical particles of mass $m$,
enclosed in a box with sides $a_{1}$, $a_{2}$ and 
$a_{3}$, at temperature $T$. We will assume a gas temperature (density) 
high (low) enough so that the Boltzmann statistics is applicable. In those
conditions, the partition function is given by$^{4}$
\be
Z = [ z ]^{N} / N! \label{eq:1}
\ee
with
\be
z = \prod_{i=1}^{3} \sum_{n_{i}=1}^{\infty} \exp(-\sigma_{i} n_{i}^{2}) 
                                                        \label{eq:2}
\ee
and
\be
\sigma_{i} = {3\over{8}}\left( {\lambda(T)\over{a_{i}}} \right)^2 \label{eq:3}
\ee
where $\lambda(T)=h/\sqrt{3 m k_{B} T}$ is the de Broglie thermal
wavelength and $k_{B}$, $h$ are the Boltzmann and Planck constants, 
respectively.
At room temperature, the 
spacing between two consecutive arguments in the exponential is very 
small (of the order of $10^{-10}$ for $m=10^{-22}\ g$, $a_{i}=10\ cm$
and $T=300\ K$). However, since $\sigma_{i}$ depends on the 
inverse of the product $temperature\ \times\  
(linear\ size\ of\ the\ system)^{2}$, 
according to Eq.(\ref{eq:3}), there are situations, such as small 
volumes (``molecular cavities'') where such 
spacing, while still small, can grow considerably (of the order of 
$10^{-3}$ for $T=1\ K$ and $a_{i}=10^{-5}\ cm$). 
 
The function we are interested in is the infinite sum
\be
s(\sigma) = \sum_{n=1}^{\infty} \exp(-\sigma n^{2}) \label{eq:4}
\ee
which is a special case of the more general type of sum  
$\sum_{n=1}^{\infty}q^{n^{2}}$ with $|q|<1$. This sum  
can be expressed in terms of the {\em theta function}$^{5}$
\be
\theta_{3}(0,q) = 1+2 q + 2 q^{4} + 2 q^{9}+...\label{eq:4p}
\ee
From the identity$^{6}$ $ (2 K/\pi)^{1/2} = \theta_{3}(0,q)$, we have
\be
s(\sigma) = {1\over{2}}\left[ \left( {2 K(k)\over{\pi}}\right)^{1/2}
                -1\right]       \label{eq:5}
\ee
where $K(k)$ is the complete elliptic integral of the first kind\\ 
$K(k) = \int_{0}^{\pi/2} d\alpha (1-k \sin(\alpha)^{2})^{-1/2}$ and the
argument $k$ satisfies the transcendental equation$^{7}$
\be
K(1-k) - (\sigma/\pi) K(k) = 0       \label{eq:6}
\ee
Fig.1 shows the plot of $K(1-k)$ and $(\sigma/\pi) K(k)$ 
for different values of $(\sigma/\pi)$. As soon as $(\sigma/\pi)$ deviates   
slightly from unity, the intersection of the two curves tends to occur very
close to either $k=0$ or $k=1$, which justifies a perturbation expansion
around those points.

At high temperatures (or large container dimensions) $(\sigma /\pi)<<1$ and 
the root of eq.(\ref{eq:6}) lies close to $k=1$,
which implies, according to eq.(\ref{eq:6}),  
$K(k)\approx (\pi/\sigma) K(0) = \pi^{2}/2\sigma$. After inserting this into
eq.(\ref{eq:5}), we obtain
\be
s(\sigma)\approx{1\over{2}}\left[ \left( {\pi\over{\sigma}}\right)^{1/2}
                -1\right]       \label{eq:7}
\ee
This approximation differs slightly from the standard textbook$^{8}$
expression $s(\sigma)=(1/2) (\pi/\sigma)^{1/2}$.
Table I shows a 
comparison of both expressions with the exact result. We see that 
eq.(\ref{eq:7}) shows better agreement up to values of $\sigma$ as high 
as $\sigma\sim1$. After inserting Eq.(\ref{eq:7}) into Eqs.(\ref{eq:2}) 
and (\ref{eq:1}), 
we have, 
\begin{eqnarray}
\log(Z) & \approx & N\log\left[ {1\over{8}}\left( {\pi^{3}\over{\sigma_{1} \sigma_{2} 
\sigma_{3}}} \right)^{1/2} {e\over{N}}\right] + 
N \sum_{i=1}^{3} \log\left(1-\left({\sigma_{i}\over{\pi}}\right)^{1/2} \right) \nonumber \\
  & \approx & N\log\left[ {1\over{8}}\left( {\pi^{3}\over{\sigma_{1} \sigma_{2} 
        \sigma_{3}}}\right)^{1/2} {e\over{N}}\right] - 
        N\sum_{i=1}^{3} \left({\sigma_{i}\over{\pi}}\right)^{1/2} 
\end{eqnarray}        
                     
We are now in position to evaluate the main corrections to the energy 
$U=k_{B}T^{2} 
(\partial \log(Z)/\partial T)_{N, V}$, pressure $P=k_{B}T (\partial \log(Z)/\partial V)_{N, T}$, 
specific heat $C=(\partial U/\partial T)_{N, V}$ and entropy $S=k_{B}\log(Z)+ 
k_{B}T(\partial \log(Z)/\partial T)_{N, V}$. We obtain : 
\be
{\mbox{Energy:}}\hspace{1.5cm}U\approx U_{0} \left[1 + {1\over{\sqrt{96 \pi}}}
\left({\lambda(T) A\over{V}}\right) \right]
\ee
where $U_{0}=(3/2) N k_{B} T$ and $A=2 (a_{1} a_{2}+a_{1} a_{3}+a_{2} a_{3})$ 
is the area of the box and $V= a_{1} a_{2} a_{3}$ is its volume.
\be
{\mbox{Pressure :}}\hspace{1.5cm}P\approx P_{0} \left[1 + 
{1\over{\sqrt{96 \pi}}}
\left({\lambda(T) A\over{V}}\right) \right]
\ee
where $P_{0}=N k_{B} T/V$.
\be
{\mbox{Specific Heat:}}\hspace{1.5cm}C\approx C_{0} \left[1 + 
{1\over{\sqrt{384 \pi}}}\left({\lambda(T) A\over{V}}\right) \right]
\ee
where $C_{0}=(3/2)N k_{B}$.
\be
{\mbox{Entropy:}}\hspace{1.5cm}S\approx S_{0} - \sqrt{{3\over{128 \pi}}}
N k_{B}\left({\lambda(T) A\over{V}}\right)
\ee
where 
\be
S_{0}=(3/2) N k_{B} + N k_{B}\log\left[ {1\over{8}}\left( {\pi^{3}\over{\sigma_{1} \sigma_{2} 
\sigma_{3}}} \right)^{1/2} {e\over{N}}\right].
\ee
Thus, the main correction due to the finite volume of the 
container has the form (thermal wavelength$\times$ area/volume).

\noindent{\bf Ideal gas in spherical container}: Let us now consider the
same ideal gas of the preceding section enclosed in a spherical
container of radius $R$. The eigenvalue equation is : $^{9}$
$j_{l}(k R) = 0$ with $l=0,1,2,...$ where $k=(8 \pi^{2} m E/h^{2})^{1/2}$
and $j_{l}(x)$ is a spherical Bessel function. For each $l$, we have 
an infinite number of solutions, indexed by an integer $n=1,2,...$ and 
for each $(l,n)$ we have an angular momentum degeneracy of $ (2 l+1)$.
The energies are $E_{l n}=(h^{2}/8\pi^{2} m R^{2}) X_{l n}^{2}$, where
$X_{l n}$ is the nth root of $j_{l}(x)$. In the limit of large R, we can 
approximate the $X_{l n}$ using McMahon's expansion: $^{10}$
\begin{eqnarray}
X_{l n} & = & b - {(\mu -1)\over{8 b}}- {4 (\mu -1) (7 \mu -31)\over{
3\ (8 b)^{3}}}-{32 (\mu -1)(83 \mu^{2}-982 \mu + 3779)\over{15\ (8 b)^{5}}}
\nonumber \\
        &   & -{64\ (\mu -1) (6949 \mu^{3}-153855 \mu^{2}+1585743 \mu -6277237)\over{
105\ (8 b)^{7}}} + ...  \label{eq:xln}
\end{eqnarray}
with $\mu\equiv (2 l+1)^{2}$ and $b\equiv (n+(l/2)) \pi$.

Now, instead of using Eq.(\ref{eq:xln}) to calculate directly the atomic 
partition function, we will use it to evaluate the {\em density of
states}, which is all we need to calculate the thermodynamic properties 
of the system. For a container of ``simple'' shape it is 
natural to expect that the main contributions to the density of states 
will have the form:
\be 
\rho(E) \approx \rho_{1} E^{1/2} + \rho_{2} + \rho_{3} E^{-1/2}  
        \label{eq:dos1}
\ee
where each term has the typical form of a three, two, and one  dimensional
density of states. The relative importance of each term in Eq.(\ref{eq:dos1}) 
will depend upon the ``dimensionality'' of the container. Thus, for 
our spherical container where ``volume'' predominates over ``area'' and
``length'', the main contribution comes from the term proportional to
$E^{1/2}$, with a small (constant) correction term. In this case, it is 
convenient to rewrite Eq.(\ref{eq:dos1}) in terms of dimensionless parameters
$R_{1},R_{2}$:
\be
\rho(E) \approx (3/2) R_{1} \left( {h^{2}\over{8 \pi^{2} m R^{2}}}\right)^{-3/2}
E^{1/2} + R_{2} \left( {h^{2}\over{8 \pi^{2} m R^{2}}}\right)^{-1}
        \label{eq:dos2}
\ee
where we have kept only the surface correction term. We proceed as follows:  
Starting from Eq.(\ref{eq:xln}), we evaluate
numerically the integrated density of states $\phi(E)$, defined 
as the number of states with energies less than $E$. Then, we perform
a least-squares fit of the resulting histogram to the smooth function
\be
\phi(E) = R_{1} \left( {h^{2}\over{8 \pi^{2} m R^{2}}}\right)^{-3/2}
E^{3/2} + R_{2} \left( {h^{2}\over{8 \pi^{2} m R^{2}}}\right)^{-1} E
\ee
obtaining the optimal values for $R_{1}, R_{2}$. The approximate 
density of states is then given by Eq.(\ref{eq:dos2}).

The atomic partition function is given by
\be
z = \int_{0}^{\infty} d E\ \  \rho(E) e^{-\beta E} .
        \label{eq:z}
\ee

By using Eqs.(\ref{eq:dos2}),(\ref{eq:z}) and the numerical values 
obtained for
$R_{1}, R_{2}$, we can write the main corrections to several
thermodynamics quantities:

\be
{\mbox{Energy:}}\hspace{1.5cm}U\approx U_{0} \left[1 + 
                0.0461 \left({\lambda(T) A\over{V}}\right) \right]
\ee
where $U_{0}=(3/2) N k_{B} T$ and $A,V$ are the area and volume of the
sphere, respectively.

\be
{\mbox{Pressure :}}\hspace{1.5cm}P\approx P_{0} \left[1 + 
                0.0461 \left({\lambda(T) A\over{V}}\right) \right]
\ee
where $P_{0}=N k_{B} T/V$.

\be
{\mbox{Specific Heat:}}\hspace{1.5cm}C\approx C_{0} \left[1 + 
                0.02305 \left({\lambda(T) A\over{V}}\right) \right]
\ee
where $C_{0}=(3/2)N k_{B}$.

\be
{\mbox{Entropy:}}\hspace{1.5cm}S\approx S_{0} - 
                0.0693 N k_{B}\left({\lambda(T) A\over{V}}\right)
\ee
where 
\be
S_{0}=(3/2) N k_{B} + N k_{B}\log\left[ 
2.804 \left( {8 m R^{2} k_{B} T\over{h^{2}}} \right)^{3/2}
{e\over{N}}\right].
\ee

The method outlined above for the sphere can be used for any other 
container of ``simple'' shape, provided the eigenergies are known either 
analytically or numerically. To its conceptual simplicity we have to oppose 
its rather slow numerical convergence, requiring very many energies to obtain 
coefficients with a given accuracy.

As in the case of the box, the main finite volume corrections 
for the sphere have the form $constant$ $\times$ 
(thermal wavelength$\times$ area/volume). 
For both, the box and the spherical container, the thermodynamical
quantities $U_{0},P_{0},C_{0}$ and $S_{0}$ do not depend on the shape
of the container. The main finite size corrections have the {\em same} 
functional dependence on the ratio (thermal wavelength$\times$ area/volume) 
for both shapes.

\vspace{1in}
%%%%%%%%%%%%%%%%%%%%%%%%%%%%%%%%%%%%%%%%%%%%%%%%%%%%%%%%%%%%%%%%%%%%%%
\noindent{\bf ACKNOWLEDGMENTS}

\noindent
I wish to thank Professor J. R\"{o}ssler and Professor R. Tabensky
for very stimulating discussions.

\newpage
%%%%%%%%%%%%%%%%%%%%%%%%%%%%%%%%%%%%%%%%%%%%%%%%%%%%%%%%%%%%%%%%%%%%%%
%\begin{thebibliography}{99}

\noindent
$^{1}$ See, for instance, Donald A. McQuarrie, {\em Statistical Mechanics}, 
(Harper and Row, New York, 1976), p. 82.

\noindent
$^{2}$ C. Stutz, ``On the Validity of Converting Sums to Integrals in
Quantum Statistical Mechanics'', Am. J. Phys. {\bf 36}, 826-829 (1968).

\noindent
$^{3}$ K. Fox, ``Comment on: `On the Validity of Converting Sums to
Integrals in Quantum Statistical Mechanics' '', Am. J. Phys. {\bf 39},
116-117 (1971).

\noindent
$^{4}$ Reference 1, p. 81.

\noindent
$^{5}$ M. Abramowitz and I.A. Stegun, {\em Handbook of Mathematical Functions},
(Dover, New York, 1965), p.576.

\noindent
$^{6}$ Reference 5, p. 579.

\noindent
$^{7}$ Reference 5, p. 591.

\noindent
$^{8}$ Reference 1, p. 82

\noindent
$^{9}$ L.D. Landau and E.M. Lifshitz, {\em Quantum Mechanics} (Pergamon, Oxford, 
New York, 1977), 3rd ed., p. 111.

\noindent
$^{10}$ Reference 5, p. 371.

%\end{thebibliography}
%%%%%%%%%%%%%%%%%%%%%%%%%%%%%%%%%%%%%%%%%%%%%%%%%%%%%%%%%%%%%%%%%%%%%
\newpage
%%%%%%%%%%%%%%%%%%%%%%%%%%%%%%%%%%%%%%%%%%%%%%%%%%%%%%%%%%%%%%%%%%%%%
\centerline{{\bf Captions List}}

\vspace{2cm}
\noindent {\bf Fig.1 :}\ \ Complete elliptic integral of the first kind. 
The continuous line is $K(1-k)$, while the dashed, 
dotted and dot-dashed curves are $(\sigma/\pi) K(k)$ for
$(\sigma/\pi)=0.5, 1$ and $2$, respectively.

\vspace{1.0cm}
\noindent {\bf Table I}\ \ Particle in a box: Comparison of the exact atomic 
partition function $s(\sigma)$ (second column) with approximation 
(\ref{eq:7}) (third column) and with the standard approximation for 
high temperatures (fourth column).

\newpage

\begin{tabular}{l c c c}\hline\hline
           &                &                & \\
\hspace{0.5cm}$\sigma$\hspace{1.0cm}   & \hspace{1cm}$\sum_{n=1}^{\infty} e^{-\sigma n^{2}}$\hspace{1cm}  & \hspace{1cm}${1\over{2}}\left[
\left({\pi\over{\sigma}}\right)^{1/2} -1\right]$\hspace{1cm} & \hspace{1cm}${1\over{2}}\left({\pi\over{\sigma}}\right)^{1/2}$\hspace{1cm} \\ 
           &                &                &               \\
\hline
     $1$    &  $0.38631860$  &  $0.38622693$  &  $0.88622693$  \\
   $0.5$    &  $0.75331414$  &  $0.75331413$  &  $1.25331414$  \\
   $0.1$    &  $2.30249561$  &  $2.30249561$  &  $2.80024956$  \\
   $0.01$   &  $8.36226926$  &  $8.36226926$  &  $8.86226926$  \\
   $0.001$  &  $27.5249561$  &  $27.5249561$  &  $28.0249561$  \\ 
   $0.0001$ &  $88.1226926$  &  $88.1226926$  &  $88.6226926$  \\
   $0.00001$&  $279.749561$  &  $279.749561$  &  $280.249561$  \\ \hline\hline
\end{tabular}
%%%%%%%%%%%%%%%%%%%%%%%%%%%%%%%%%%%%%%%%%%%%%%%%%%%%%%%%%%%%%%%%%%%%%%%%%%
\newpage
%%%%%%%%%%%%% FIG. 1 goes here %%%%%%%%%%%%%%%%%%%%%%%%%%%%%%%%%%%%%
%
\begin{figure}[p]
\begin{center}
\leavevmode
\hbox{
\includegraphics{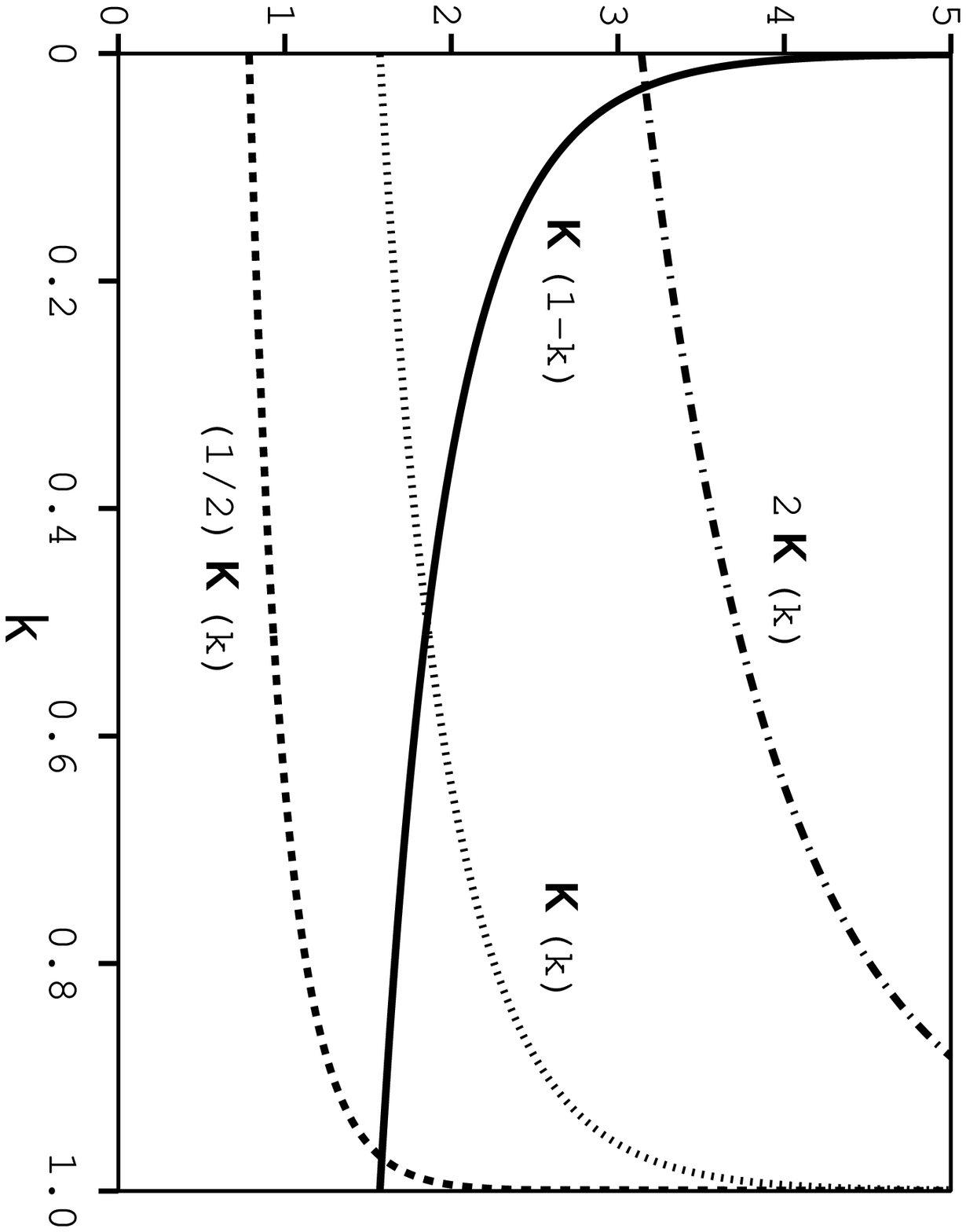}  }	
%\special{psfile=potentia.eps hoffset=200 voffset=-270
%hscale=55 vscale=55 angle=90}  }
\end{center}
\vspace{2.3 in}
\label{figure1}
\vspace{4.0cm}
\end{figure}

\end{document}